\documentstyle[aps,prb,twoside,multicol,epsfig]{revtex}
\begin{document}

\title{On the unitarity of S-matrix in 1-d case}

\author{                  Dimitar I. Pushkarov}

\address{
                 Institute of Solid State Physics
      Bulgarian Academy of Sciences,
       Sofia 1784, Bulgaria,\\
       E-mail: dipushk@issp.bas.bg
 }
\maketitle
\begin{abstract}
       It is shown that the scattering S-matrix is unitary
even if the scattering potential $U(x)$ tends to different limits
at $x \rightarrow\pm \infty$. This result is in contrast to the
statements of some authors which argue that the unitarity may be
broken in 1-d case if $U(-\infty) \ne U(+\infty)$. The mistake may
result from a wrong normalization of the wave functions. This work
may be considered as a comment to some of those works.
\end{abstract}

\begin{multicols}{2}
 A fundamental property of the scattering S-matrix is
its unitarity (c.f. e.g. Ref.[1]). This is a consequence of the
particle conservation law. However, there have appeared a number
of works \cite{K1,K2,K3,K4} where the S-matrix obtained or used is
not unitary, and the authors have accepted this as a special
feature of the problem owing to the different behavior of the
potential at $\pm \infty$. This work may be considered as a
comment on those works. It consists of two parts. First, we repeat
a wrong derivation using a simple example and then prove the
unitarity of the scattering matrix in the general case.

Let us consider a step potential of the form:

\begin{eqnarray}\label{1}
U(x) =\left\{
\begin{array}{cc}
0 \ , \quad  x <  0 &
 \\
U_0, \quad x > 0
\end{array}
\right.
\end{eqnarray}
The general solution of the Schr\"odinger equation is well known:
\begin{eqnarray}\label{2}
  \psi_l = A e^{ikx} + B e^{-ikx}, \qquad x<0 \nonumber\\
  \psi_r = C e^{i\kappa x} + D e^{-\kappa x}, \qquad x>0
\end{eqnarray}
where $ k = \sqrt{2E}$ \ ,  $\kappa = \sqrt {2(E-U_0)}$ \ and for
simplicity we put $\hbar= m = 1$. Let us consider the most
instructive case when $E>U_0$. The constants $A,B,C,D$ have to
satisfy the relations $\psi_l(0) = \psi_r(0)$ and $\psi'_l(0) =
\psi'_r(0)$. This yields
\begin{equation}\label{3}
  A + B = C + D, \qquad k(A-B) = \kappa (C-D)
\end{equation}

The S-matrix components connect the incoming waves ($ A e^{ikx}$
and $De^{-\kappa x}$) with the outgoing ones ($B e^{-ikx}$ and $C
e^{i\kappa x}$). If one defines the S-matrix by the relation
\begin{eqnarray}\label{4}
\left(
\begin{array}{cc}
A
 \\
D
\end{array}
\right) = \tilde S \left(
\begin{array}{cc} C
 \\ B
\end{array}
\right)
\end{eqnarray}
it will be of the form
\begin{equation}\label{5}
\tilde S_{11}=\frac{2\kappa}{k + \kappa}, \,\, \tilde S_{22}=
\frac{2k}{k + \kappa}, \,\, \tilde S_{12} = -\tilde S_{21} =
\frac{k-\kappa}{k + \kappa}
\end{equation}
The quantity $R = \tilde S^2_{12} = \tilde S^2_{21}$ coincides
with the reflection coefficient, but $\tilde S_{11}$ and $\tilde
S_{22}$ have no physical meaning. They are not transmission
amplitudes, at least because ${\tilde S_{11}}^2 + {\tilde
S_{12}}^2 \ne 1$. \textit{Therefore, $\tilde S$ is not the
scattering matrix of our problem.}

    Let us now derive the correct form of the S-matrix. We shall do this
for an arbitrary potential of the form:
\begin{equation}\label{6}
  U(x\rightarrow -\infty)= 0
  , \quad
  U(x\rightarrow \infty) = U_0
\end{equation}
Therefore, the asymptotes at $\pm \infty$ are of the form
(\ref{2}). However, for calculating the scattering S-matrix
components \textit{these functions have to be normalized to unit
flux, not to unit density}! Only in this case the S-matrix
components correspond to reflection and transmission amplitudes
(the corresponding coefficients are ratios of particle fluxes).
So, we write, instead of (\ref{2}), the following functions:
\begin{equation}\label{7}
\psi_L = A_1 \frac{ e^{ikx}}{\sqrt k} + B_1 \frac{e^{-ikx}} {\sqrt
k}, \qquad x\rightarrow   -\infty
\end{equation}
\begin{equation}\label{8}
  \psi_R = A_2 \frac{e^{i\kappa x}}{\sqrt \kappa} +
   B_2 \frac{e^{-\kappa x}}{\sqrt \kappa},
  \qquad x\rightarrow \infty
\end{equation}
In (7) and (8) $k$ and $\kappa$ coincide with the particle
velocities at infinity because we put $\hbar=m=1$. The functions
$\psi_L$ and $\psi_R$ are asymptotes of one and the same exact
solution. Hence, the following relations between coefficients
$A_i$ and $B_i$ exist \cite{LLQ}:
\begin{equation}\label{9}
  A_2 = \alpha A_1 + \beta B_1, \quad B_2 = \beta^* A_1 + \alpha^*
  B_1
\end{equation}
Since the wave functions are now normalized to unit fluxes, the
particle conservation law (or the continuity condition for the
current) reads
\begin{equation}\label{10}
|A_1|^2 - |B_1|^2 = |A_2|^2 - |B_2|^2
\end{equation}
Substituting here $A_2$ and $B_2$ from (9) leads to
\begin{equation}\label{11}
  |\alpha|^2 - |\beta|^2 = 1
\end{equation}
One has from (9):
\begin{equation}\label{12}
  A_1 = \frac{1}{\alpha}A_2 - \frac{\beta}{\alpha}B_1, \quad B_2 =
  \frac{\beta^*}{\alpha}A_2 + \frac{1}{\alpha}B_1
\end{equation}
This means, that the scattering matrix has the form:
\begin{eqnarray}\label{13}
\hat S = \left(
\begin{array}{cc} 1/\alpha  & -{\beta}/{\alpha}
 \\\\
 {\beta^*}/{\alpha}          &  {1}/{\alpha}
\end{array}
\right)
\end{eqnarray}
It is easy to check using (11), that this matrix is unitary. The
reflection and transmission coefficients are respectively
$$
R = |\beta/\alpha|^2 = |\beta/\alpha^*|^2, \quad T = |\alpha|^{-2}
$$
and $T + R = 1$ as it should be.

     It is instructive to note in conclusion, that the problem of
     normalization does not appear in the typical 3-d situations
     where the scattering potential vanishes at infinity.
     Nevertheless it is helpful sometimes to use the normalization
     to unit flux as e.g. in the Born approximation (c.f. Ref.
     [1] \S\ 126).


\end{multicols}
\end{document}